\documentclass{river-journal}
\usepackage{fancyhdr}
\usepackage{rivps}
\usepackage{fancyhdr}
\usepackage{graphicx}
\usepackage{graphics}
\usepackage{cite}
\usepackage{amsmath}
\usepackage{amsfonts}
\usepackage{gensymb}
\usepackage{amssymb}
\usepackage{amsmath}
\usepackage{algorithm,algorithmic}
\usepackage{tikz}
\usetikzlibrary{arrows}
\usepackage{pgfplots}
\usepackage{textcomp}
\usepackage{multirow}
\usepackage{siunitx}
\usepackage[mathscr]{euscript}
\usepackage{xspace}
\usepackage{enumerate}

\usepackage{subcaption}
\usepackage{hyperref}
\usepackage{url}
\usepackage{multirow}

\usepackage{algorithm,algorithmic}
\usepackage{tikz}
\usepackage{pgfplots}
\usepackage{textcomp}
\usepackage[utf8]{inputenc}
\usepackage{siunitx}
\pgfplotsset{every axis/.append style={
		label style={font=\scriptsize},
		tick label style={font=\scriptsize},
		legend style={font=\scriptsize}
}}

\par
	
\begin{document}
	\begin{center}
	This article has been accepted in Journal of Web Engineering © 2021 River Publisher. Personal use of this material is permitted. Permission from River Publisher	must be obtained for all other uses, in any current or future media, including reprinting/republishing this material for advertising or promotional purposes, creating new collective works, for resale or redistribution to servers or lists, or reuse of any copyrighted component of this work in other works.
		\smallskip
		\hrule 
		\smallskip

	\end{center}
	
	\begin{center}
		\bfseries Dynamic Resource Allocation Method for Load Balance Scheduling over Cloud Data Center Networks
		
		\bigskip
		\textbf{Sakshi Chhabra, Ashutosh Kumar Singh}\\
			National Institute of Technology Kurukshetra 
			Haryana, INDIA\\
			sakshichhabra555@gmail.com\\
			ashutosh@nitkkr.ac.in
		
	\end{center}
	
		\title{Dynamic Resource Allocation Method for Load Balance Scheduling over Cloud Data Center Networks}
		\author{Sakshi Chhabra, Ashutosh Kumar Singh}
		\institute{Sakshi Chhabra,
			National Institute of Technology Kurukshetra 
			Haryana, INDIA\\
			\email{sakshichhabra555@gmail.com},\\
			Ashutosh Kumar Singh,
			National Institute of Technology Kurukshetra 
			Haryana, INDIA
			\email{ashutosh@nitkkr.ac.in}}
	
	\runningtitle{Resource Allocation Method for Load Balance Scheduling}
	\runningauthor{Sakshi Chhabra et. al.}
	
	\subsection*{Abstract}
	The cloud datacenter has numerous hosts as well as application requests where resources are dynamic. The
	demands placed on the resource allocation
	are diverse. These factors could lead to load imbalances, which
	affect scheduling efficiency and resource utilization. A scheduling
	method called Dynamic Resource Allocation for Load Balancing (DRALB) is proposed. The proposed solution constitutes two steps: First, the load manager analyzes the resource requirements such as CPU, Memory, Energy and Bandwidth usage and allocates an appropriate number of VMs for each application. Second, the resource information is collected and updated where resources are sorted
	into four queues according to the loads of resources i.e. CPU intensive, Memory intensive, Energy intensive and Bandwidth intensive. We demonstarate that SLA-aware scheduling not only facilitates the cloud consumers by resources availability and improves throughput, response time etc. but also maximizes the cloud profits with less resource utilization and SLA (Service Level Agreement) violation penalties. This method is based on diversity of client's applications and searching the optimal resources for the particular deployment. Experiments were carried out based on following parameters i.e. average
	response time; resource utilization, SLA violation rate and load balancing. The experimental results demonstrate that this
	method can reduce the wastage of resources and reduces the traffic upto 44.89\% and 58.49\% in the network.\\	
	\section{Introduction}
	The rapid development of cloud computing has increased the traffic rate exponentially in the data center networks. In order to save storage, energy consumption, bandwidth and computing capacity we need efficient cloud traffic engineering within each data datacenter \cite{1}. Therefore, how to handle such a growing amount of workload in a scalable manner and optimized the Virtual Machine (VM) Placement to accommodate the traffic growth. To maximize the resource utilization and satisfying the Service Level Agreement (SLA) for tenants are important research issues. It requires effective management of cloud resource provisioning. Cloud provisioning is the allocation of resources to the clients according to their requirements, some require powerful CPU computing capacity or high bandwidth or large amount of storage which leads the load imbalance problem. In our daily scenario, clients could experience many difficulties like long system immediate responses during bank deposits or withdrawls, real-time temperature measurements, delays etc. For these application tasks, the real-time resource manager and load balancer signify the decision about the computing resources and the load estimation \cite{2}. Hence, task scheduling and resource management play a key role in cloud computing to maximize the diversity in client's applications and the uncertain factors of resources. On the other hand, Service Level Agreement (SLA) is a part of service contract and one of the major considerations for every cloud's customer. However, it becomes challenging for cloud providers to meet SLA due to dynamic multiresource sharing. For some sensitive applications, a minimum of 99.9\% availability is required which difficult for various cloud computing services \cite{3}. In this work, we aim to provide SLA aware cloud resource provision framework which ensures the Quality of Service (QOS) with least violation rate. We demonstarate that SLA-aware scheduling not only facilitates the cloud consumers by resources availability and improves throughput, response time etc. but also maximizes the cloud profit with less resources utilization and SLA violation penalty. The paper aims to propose a dynamic resource allocation method called DRALB for scheduling the workload over cloud data cenetrs. This method is based on diversity of client's applications and searching the optimal resources for the particular deployment.
	
	The rest of the paper is organized with following sections: Section 2 introduces the related work. Section 3 shows a proposed framework for cloud data centers. Section 4 includes performance evaluation and analysis of implementing the proposed work followed by conclusion in section 5.
	\section{Related Work}
	To study the optimal resource scheduling during task deployment in cloud computing, several techniques have been proposed by various authors and a few of them are explained. Fung Po Tso et. al. \cite{4} have discovered a
	technique of improving data center utilization based on two network topologies:
	canonical and fat tree. This model was found effective to improve
	utilization by using near optimal traffic engineering and it reduces Maximum Link Utilization (MLU)
	and increase overall network capacity through Penalizing Exponential Flow-spliTing (PEFT) routing. Liang Quan et. al. \cite{5} have presented a reconfiguration framework based on request predication that determines the objective of relatively optimal configuration. They have evaluated their algorithm with request prediction and deal with App\_VM\_Configuration, Assignment Shifting and Deployment Shifting. An optimal VM Placement method for traffic scalability have been explained in \cite{6} whch formulates the idea of Marcov Chain (MC) based solution to optimize the VM placements. This algorithm decreases the rate of exchanging traffic among racks and avoids the traffic overflow. Liyun Zuo et. al. \cite{7} have proposed a multique interlacing method based on task's classification where resources are sorted into three queues: CPU, I/O, Memory-intensive according to their task's requirements. This method was found effective to balance the load that were added to improve the resource utilization and performance. According to the evolved results, it was found that their algorithm is always better than previous solutions especially for large number of tasks. This application type based VM placement and allocation strategy is proposed in \cite{8}. By comparing and analyzing the resource usage efficiency and improves the application execution, which represents better performance. Experimental results menifest the two strategies i.e. VMAllocationPolicy (VAPS) and LoadBalanceVMAllocation. The proposed method in \cite{9} are implemented in Cloudsim have presented a dynamic hierarchical load balancing model which helps to solves the traffic scalability issue. This framework selects the most approporiate host that satisfies the multi-dimensional resource constraints over random and sequential. DHLB improves upto 66\% and outperforms the existing solutions. Wanchun Dou et. al. \cite{10} have designed a hierarchical control framework for leveraging task scheduling and Resource Co-Allocation (RCA) method for the big data platform. This framework consists of four steps: 1) Meta service preprocessing 2) Resource usage monitoring 3) Resource co-allocation for meta services 4) Global resouce co-allocation. The cloud computing benefits cloud service consumers in terms of cost and helps to reduce temporal and monetary costs. By evaluation and analyzed the performance of cost optimization parameters is explained in \cite{11}. The dynamic hyper-heuristic technique that can effectively optimize and save the cost and time of cloud service providers has been provided. Completion Time Driven Hyper-Heuristic (CTDHH) has achieved the optimal results for Scientific Workflow Applications (SWFA) datasets. As cloud consumers are requesting SLAs in order to use services with acceptable QOS. Some work has \cite{12}, presented a SLA-aware resource scheduling framework i.e. dynamic hybrid metaheuristic algorithm to maximize the profits based on Parallel SA (PSA) and simulated annealing. \\
	All reseachers mentioned above worked against the resource scheduling and load balancing problems. Through different from the previous studies, we focus on a real-time application request types with resource configuration and optimization. The model allocates an appropriate resources to the VMs based on its types. Through analysis of our proposed algorithm, the VM allocation scheme is found suitable and guarantees that load among physical machines is well balanced and comparatively superior to the above works.
	
	\section{DRALB}
	In IaaS cloud data centers, when users submit the task requests, the cloud DC choose the hosts randomly to deploy the tasks. But it becomes optimal if we choose the optimal hosts for the particular task deployment. The problem of task deployment is formalized as follows: consider a set of n resources $\Re=\{r_1, r_2,\dots,r_n\}$ and $t$ task requests $T=\{T_1, T_2,\dots,T_t\}$ in the current system of cloud. The set of VMs represented by $v$ virtual machines $\rtimes=\{VM_1,VM_2,\dots,VM_v\}$ need to be placed into physical hosts $\bigotimes=\{ph_1,ph_2,\dots,ph_p\}$ formulated with $n$ clients  $\Cup= \{Cl_1, Cl_2,\dots,Cl_n\}$, $m$ server set $\circledS= \{s_1, s_2,\dots,s_m\}$. Given such scenario, the CSPs always desire to obtain an optimal mapping of VMs and servers to maximize the resource utilization. A mapping between physical host and clients with specific requests $\varTheta:T \times \Cup \rightarrow \bigotimes$ allocates each physical host from each user with specific task requests, if $\circledS$ hosts one or more VMs, it is active, $\varTheta_{T \times \bigotimes \times \Cup} =\{ \varTheta_{T_t,ph_p,Cl_n}|\varTheta_{T_t,ph_p,Cl_n}=1$ if task request $T$ of client $n$ is allocated to optimal physical host\}. Similarly, if VMs were assigned to optimal hosts then $\varTheta_i$=1 otherwise $\varTheta_i$=0. Let $\uplus_i$ represents the resources which has $\uplus_i^C$,  $\uplus_i^M$, $\uplus_i^E$ and $\uplus_i^B$ capacity of CPU, Memory, Energy and Bandwidth respectively. The utilization of resources for ${VM}_j$ are ${VM}_j^C$, ${VM}_j^M$, ${VM}_j^E$ and ${VM}_j^B$. At the time of allocating tasks, firstly we check whether available memory is greater or close to the requested ones then it can only deploy the tasks. As Eq. (1) ensures that the total required consumption of processors resource amount should not exceed its total capacity. 
	\begin{equation}
	\sum_{i=1}^{VM} {VM}_j^C \leq \uplus_i^C \,\, \wedge  \,\, {VM}_j^M \leq \uplus_i^M\,\,\, \wedge \,\, {VM}_j^E \leq \uplus_i^E \,\,\wedge \,\, {VM}_j^B \leq \uplus_i^B\,,\,\, \forall s_m \in \circledS
	\end{equation}
	The resource utilization of each resource is calculated independently, CPU utilization of one physical host is computed using Eq. (2), memory, energy and bandwidth respectively in Eq. (3-5). The average resource utilization of datacenter $\Phi_{DC}$ that needs to be maximized is demonstrated using Eq. (6). $|N|$ represents the number of resources to be considered i.e. $|N|=4$.
	\begin{equation}
	\Phi_i^C= \frac{\sum_{i=1}^{n} \varTheta_i \times {VM}_j^C }{\uplus_i^C}
	\end{equation}
	\begin{equation}
	\Phi_i^M= \frac{\sum_{i=1}^{n} \varTheta_i \times {VM}_j^M }{\uplus_i^M}
	\end{equation}
	\begin{equation}
	\Phi_i^E= \frac{\sum_{i=1}^{n} \varTheta_i \times {VM}_j^E }{\uplus_i^E}
	\end{equation}
	\begin{equation}
	\Phi_i^B= \frac{\sum_{i=1}^{n} \varTheta_i \times {VM}_j^B }{\uplus_i^B}
	\end{equation}
	\begin{equation}
	\Phi_{DC}=\frac{\sum_{i=1}^{t} \Phi_{i=1}^C + \sum_{i=1}^{t} \Phi_{i=1}^M + \sum_{i=1}^{t} \Phi_{i=1}^E + \sum_{i=1}^{t} \Phi_{i=1}^B }{|N| \times \sum_{i=1}^{t} \varTheta_i}
	\end{equation}
	\begin{figure} [!h]
		\centering
		\includegraphics[clip,trim=0cm 0cm 0cm 0cm, width=0.98\linewidth, height=10cm]{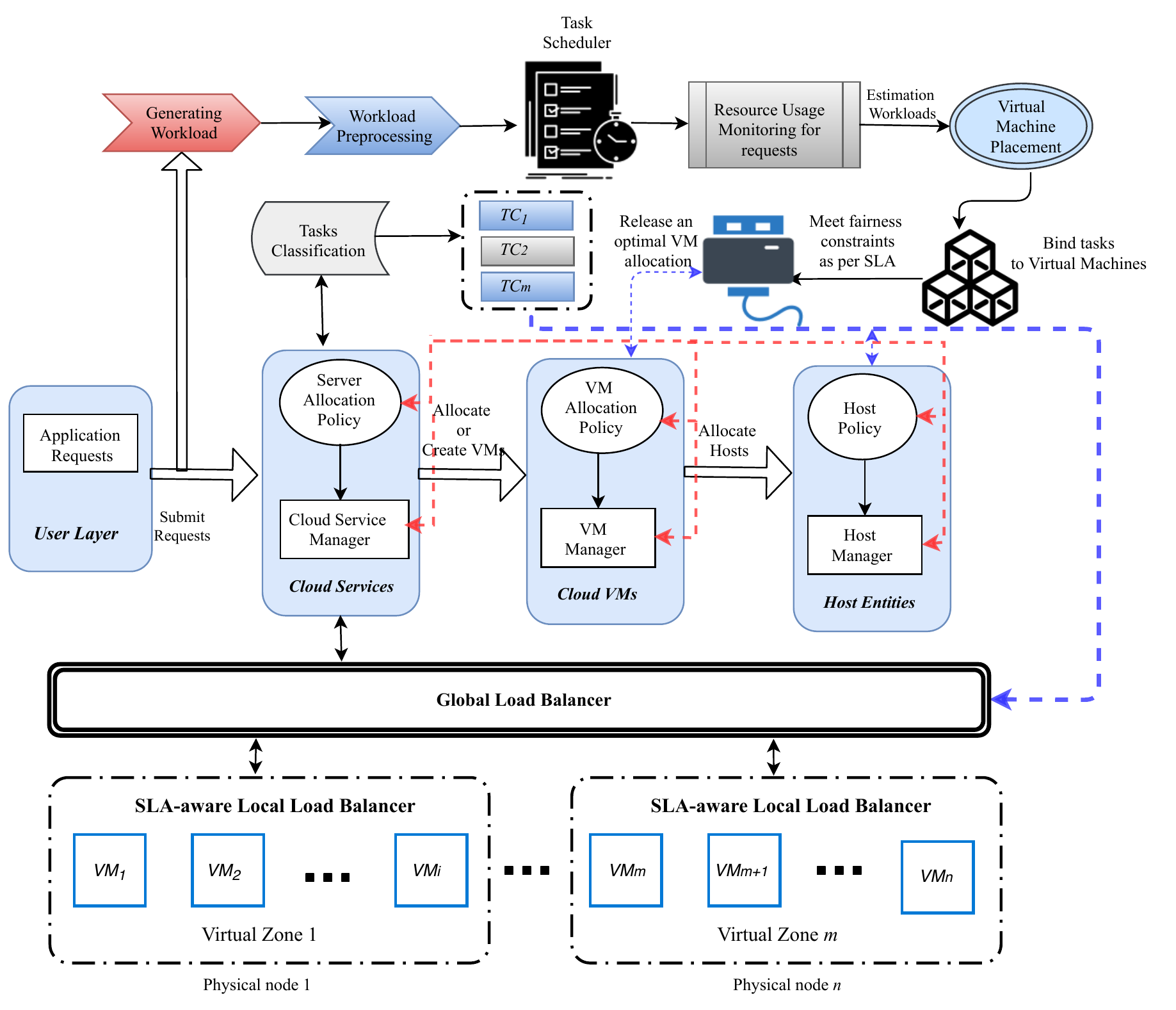}	
		\caption{DRALB Framework}
		\label{Multi-tenancy attacks during Load Balancing}
		
	\end{figure}
	
	The model chooses the more appropriate physical machine for the VM migration and placement to the respective upcoming applications. The effective VM placement optimizes the resource utilization, performance and energy consumption without SLA violation. 
	
	\begin{equation}
	f\{ph_1,ph_2,\dots,ph_n|\uplus\}=f(ph_1|\uplus)\times f(ph_2|\uplus)\times \dots \times f(ph_n|\uplus).
	\end{equation}
	where $\uplus$ are the parameters i.e. CPU, memory, energy consumption and bandwidth. The upcoming application requests are classified according to their type and resource requirements. By analyzing the resource need for the respective application, search the appropriate VM and PM where task can be deployed.  For example, if
	the weights of an unknown application $W_{ci}$ = 1, $W_{mi}$ = 0, $W_{ei}$ = 0,
	$W_{bwi}$ = 0 is obtained through an application classifier algorithm,  it means the application is CPU-intensive. Generally the usage of CPU is more for the upcoming applications as per the proposed VM allocation and placement strategy. 
	More CPU resource should be allocated to the VM, while the other
	resources such as memory, energy, bandwidth, etc. can be relatively
	less. 
	\subsection{SLA-aware Cloud Provision Model}
	This framework monitors the upcoming applications and demonstrate the SLA aware scheduling policy. The two parameters are mainly focused: Response time (RT) and Resource Utilization criteria (RUC). The response time is defined in SLA agreement to calculate the time of upcoming applications. The threshold is fixed that defines the maximum average response time to deal with tasks. As RUC threshold represents the maximum utilization of resources that every VM instance can have. If the utilization or response time exceeds the threshold number
	then apply penalty on it. 
	\begin{equation}
	PF=\sum_{i=0}^{n}[Pr_i \times \uplus_i - (Pnl_i \times U_i + Cost \times \uplus_i)]
	\end{equation}
	where $Pr_i$ defines the pricing, $Pnl_i$ is penalty. The objective of this SLA-aware cloud load balancer is to maximize the benefit of cloud services and minimizes the penalty function. The cost and penalty should be less during cloud service delivery. The two performance parameters are focused to meet fairness constraints as per SLA violation while monitoring the applications.
	Fig. 1 describes the architecture of proposed DRALB approach in the cloud environment. It shows the interaction between Task Scheduler and Resource Manager which plays an important role in the whole architecture. Resource manager monitors the link capacity and link state routing utilization for every host. We adopt task classifier
	to classify the resource requirements (CPU, memory, energy and bandwidth utilization) for each application as per SLA. When clients are requesting new batch of jobs arriving at cloud system, then load balancing is activated. The test data is monitored and initialized by load managers. Then optimal VM placement is applied and selects the most favorable machine which accomplishes and fulfill the requirement of deploying tasks. The SLA aware load balancer is divided into two types: local and global load balancer. Local load balancer monitors the load which are in same physical host to meet the SLA aware requirements. Global balancer transfer the upcoming requests to the under-loaded physical hosts as per need of SLA. Moreover, the energy consumption is calculated:
	\begin{align}
	\exists_k=
	\begin{cases}
	\Im \times \Big[\uplus_k \times \eth_k^{work} + (1-\uplus_k) \times \eth_k^{idle}\Big],Host state:on  \\
	\Im \times \eth_k^{standby},\,\,\,\,\, \, \, \, \, \, \, \,\,\,\, \, \,Host-state:\, off
	\end{cases}
	\end{align}
	The total energy consumption of a cloud data center is defined as $\exists=\sum_{k=0}^{n}\exists_k$. If the host state is on then it comes either in work state or in idle state. It is assumed that these hosts are heterogeneous and implemented in dynamic environment. 
	\section{Performance Evaluation}
	The experiments are conducted using Cloudsim 3.0 and Java-Eclipse IDE on a machine equipped with Intel\circledR\  Core\texttrademark\ I5-3230M processor of 2.60 GHZ clock speed and 8 GB of main memory to confirm the performance of DRALB. 
	\subsection{Experimental Setup}
	The simulation cloud network is carried out which were generated by Cloudsim and had clear CPU, Memory, Energy and bandwidth. The results are performed on 0 to 200 physical hosts with different configurations and had 10 virtual machines on each host. We have compared our model's performance to that of widely adopted well-known methods employed in literature, in terms of well-defined set of metrics. A series of task requests has been generated on each experiment and every task request has obvious need for computing resource amount of CPU, Memory, Energy and Bandwidth. DRALB model is compared with other three algorithms \cite{9}: random, sequential and DHLB. The number of tasks used in this experiment to verify the model's efficacy is T=400 (T$>$R) and R=1000 (T$>$R). The parameters used in simulation is illustrated in Table 1.	
	\begin{table} [!h]
		\caption{Parameters Used in Simulation}
		\begin{center}
			\begin{tabular}{ll}
				\hline 
				Parameters & Value\\
				\hline
				\textbf{VM Setup of Data Center}\\
				\hline
				CPU Computing ability & 1860 MIPs, 2660 MIPs\\
				Disk I/O & 8 GB\\
				RAM & 4096 MB\\
				Bandwidth & 100 M/s\\
				Storage & 10 G \\
				\hline
				\textbf{Task Setup of Data Center}\\
				\hline
				Length (CPU) & [250-1000] MIPs\\
				File Size & [100-2000] MB\\
				Output size (Memory) & [20-40] MB\\
				\hline
			\end{tabular}
		\end{center}
	\end{table}
	\subsection{Experimental Results}
	The following experiment first tests the makespan time which defines the total time needed for processing the tasks from begining to the end. It is mainly used in the context of scheduling when the job requests to physical hosts are assigned. The first parameter of effective load balancing is time where:
	\begin{equation}
	TT_i=\sum P_t + R_t + W_t
	\end{equation}
	where $P_t$, $R_t$ and $W_t$ are the processing time, recieving time and waiting time respectively. The second is resource utilization, DRALB minimises the wastage of resources while maximising the performance and its usage of their data centres. The third is average response time which is used to evaluate the scheduling performance. Then the load balancing verifies the load and achieves the overload avoidance for systems with multi-resource constraints. The finding of the failure nodes in the dynamic environment can only be possible if the chosen physical machine can't fulfill some of the demands of requested tasks. Then, we calculated the SLA violation rate 
	to measure the availability of services for customers in cloud systems.
	\begin{equation}
	SLA_{Vrate}= \frac{Number of Requests Violated}{Number of Total Requests}
	\end{equation}
	Table 2 illustrates the makespan which defines the total time needed for processing the tasks from begining to the end. It is mainly used in the context of scheduling when the job requests to physical hosts are assigned. Table 3 depicts the \%age of resource utilisation and DRALB minimises the wastage of resources while maximising the performance and its usage of their data centres. DRALB improves the utilisation of resources up to 38.71\% , 33.24\% and 21.98\% for random, sequential and DHLB respectively. 
	Fig. 2 shows the performance of evaluated algorithm DRALB which produces the average response time than other two heuristics when Tasks ($T$)=400 and Tasks $<$ Resources (R), as it considers both the current system state and the future VM placement request arrivals. Fig.3 evaluates the average response time when ($T$)=1000 and $T>R$ to analyze and evaluate whether it is good or bad in terms of its service performance. These effective measures mainly include the ability of dealing with tasks, the response time to calculate a task request and number of completed service per unit time etc.  The finding of these failure nodes in the dynamic environment can only be possible if the chosen physical machine can't fulfill some of the demands of requested tasks as demonstrated in Fig. 4.  Fig. 5 shows the proportion of traffic overflow with other existing techniques in the simulated cloud network data center. 
	By finding the optimal resources during load balancing, the model distributes
	client requests or network load efficiently across multiple servers that results
	in reduced traffic. During simulation, we have calculated the proportion of
	traffic that are sent to a server. The traffic percentage is calculated by Traffic
	percentage = (Assigned weight/Total weight) *100. It has been found that
	the proposed work reduced the traffic upto 58.49\% during experimental evaluation.
	It achieves the overload avoidance for systems with multi-resource constraints.
	\begin{table} 
		\caption{Makespan for Requested Tasks}\label{tbl:headings} 
		\centering
		\begin{tabular}
			{|p{1.4cm}||p{0.8cm}|p{0.8cm}|p{0.9cm}|p{1.2cm}||p{0.8cm}|p{0.8cm}|p{0.9cm}|p{1.2cm}|}
			\hline
			& \multicolumn{8}{c|}{Number of VMs} \\
			\cline{2-9}
			Requested & \multicolumn{4}{c|}{100} & \multicolumn{4}{c|}{200}\\
			\cline{2-9}
			Tasks & RND & SEQ & DHLB &DRALB	& RND & SEQ & DHLB &DRALB\\
			\hline
			\hline
				
			40  &27.786 &26.564 &18.998 &10.439 &14.532 &20.718 &8.674 &4.721 \\
			
			80 &22.742 &28.948 &21.730 &11.299 &18.529 &19.630 &10.087 &7.248 \\
			
			120 &31.589 &31.632 &23.321 &13.895 &12.630 &23.751 &11.929 &9.659\\
			
			160 &25.857 &34.704 &28.779  &15.075 &13.652 &25.859 &14.840 &12.840 \\
			
			200 &39.059 &37.736 &31.729 &16.639 &21.933 &26.259 &18.692 &16.552  \\
			\hline
		\end{tabular}
	\end{table}

	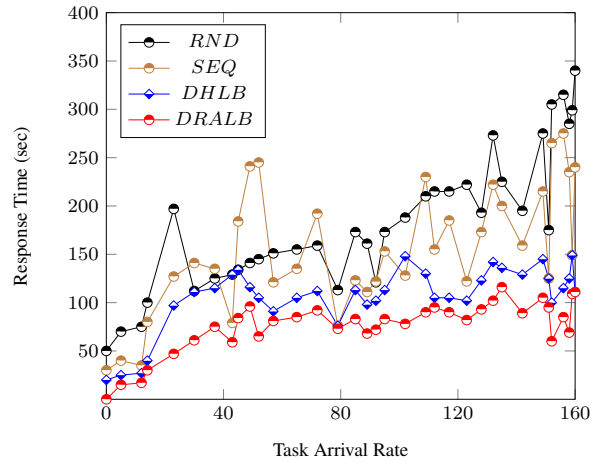
\begin{figure} [!h]
		\centering
		\begin{tikzpicture} [scale=0.9]
		\begin{axis}[
		xlabel={Task Arrival Rate},
		ylabel={Response Time (sec)},
		xmin=00, xmax=160,
		ymin=00, ymax=400,
		xtick={0,40,80,120,160},
		ytick={50,100,150,200,250,300,350,400,500,600},
		legend pos=north west,
		]
		\addplot
		[
		color=black,
		mark=halfcircle*,
		]  coordinates {(0,50) (5,70) (12,75) (14,100) (23,197) (30,112) (37,125)	(43,129)
			(45,134) (49,141) (52,145)	(57,151)
			(65,155) (72,159) (79,113) (85,173) (89,161) (92,121) (95,173) (102,188)	(109,210)
			(112,215) (117,215) (123,222) (128,193) (132,273) (135,225) (142,195) (149,275) (151,175) (152,305) (156,315) (158,285) (159,299) (160,340)	
		};
		\addplot
		[
		color=brown,
		mark=halfcircle*,
		]   coordinates {(0,30) (5,40) (12,35) (14,80) (23,127) (30,141) (37,135)	(43,79)
			(45,184) (49,241) (52,245)	(57,121)
			(65,135) (72,192) (79,76) (85,123) (89,111) (92,122) (95,153) (102,128)	(109,230)
			(112,155) (117,185) (123,122) (128,173) (132,222) (135,200) (142,159) (149,215) (151,125) (152,265) (156,275) (158,235) (159,149) (160,240)	
		};
		\addplot [
		color=blue,
		mark=halfsquare*,
		] coordinates {(0,20) (5,25) (12,27) (14,40) (23,97) (30,111) (37,115)	(43,129)
			(45,134) (49,116) (52,105)	(57,91)
			(65,105) (72,112) (79,76) (85,113) (89,98) (92,102) (95,113) (102,148)	(109,130)
			(112,105) (117,105) (123,102) (128,123) (132,142) (135,136) (142,129) (149,145) (151,125) (152,100) (156,115) (158,125) (159,149) (160,110)	
		};
		\addplot [
		color=red,
		mark=halfcircle*,
		] coordinates {(0,0) (5,15) (12,17) (14,30) (23,47) (30,61) (37,75)	(43,59)
			(45,84) (49,96) (52,65)	(57,81)
			(65,85) (72,92) (79,73) (85,83) (89,68) (92,72) (95,83) (102,78)	(109,90)
			(112,95) (117,90) (123,82) (128,93) (132,102) (135,116) (142,89) (149,105) (151,95) (152,60) (156,85) (158,69) (159,109) (160,111)	
		};
		\legend{$RND$,$SEQ$,$DHLB$,$DRALB$}
		\end{axis}
		
		\end{tikzpicture}
		\caption{Comparison of RND, SEQ, DHLB and DRALB in terms of Average Response Time when T=400 and T$<$R}
		\label{Comparison of Time-Shared and Space-Shared resource allocation in terms of Failures}

		
	\end{figure}

	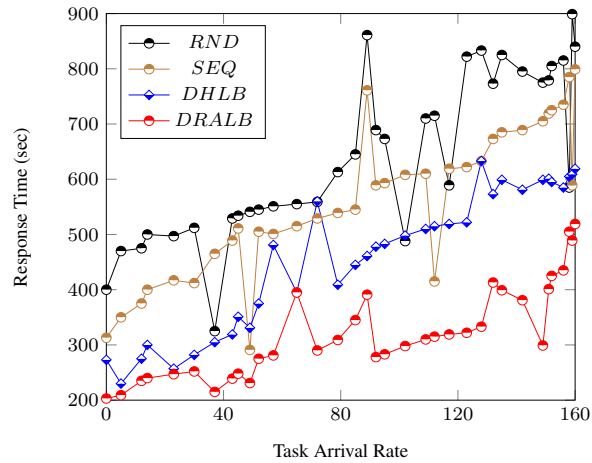
\begin{figure} [!h]
		\centering
		\begin{tikzpicture} [scale=0.9]
		\begin{axis}[
		xlabel={Task Arrival Rate},
		ylabel={Response Time (sec)},
		xmin=00, xmax=160,
		ymin=200, ymax=900,
		xtick={0,40,80,120,160},
		ytick={200,300,400,500,600,700,800,900},
		legend pos=north west,
		]
		\addplot
		[
		color=black,
		mark=halfcircle*,
		]  coordinates {(0,400) (5,470) (12,475) (14,500) (23,497) (30,512) (37,325)	(43,529)
			(45,534) (49,541) (52,545)	(57,551)
			(65,555) (72,559) (79,613) (85,645) (89,861) (92,689) (95,673) (102,488)	(109,710)
			(112,715) (117,589) (123,822) (128,833) (132,773) (135,825) (142,795) (149,775) (151,779) (152,805) (156,815) (158,585) (159,899) (160,840)	
		};
		\addplot
		[
		color=brown,
		mark=halfcircle*,
		]   coordinates {(0,313) (5,350) (12,375) (14,400) (23,417) (30,412) (37,465)	(43,489)
			(45,511) (49,291) (52,505)	(57,501)
			(65,515) (72,529) (79,539) (85,545) (89,761) (92,589) (95,593) (102,608)	(109,610)
			(112,415) (117,619) (123,622) (128,633) (132,673) (135,685) (142,689) (149,705) (151,719) (152,725) (156,735) (158,785) (159,589) (160,799)	
		};
		\addplot [
		color=blue,
		mark=halfsquare*,
		] coordinates {(0,273) (5,230) (12,275) (14,300) (23,257) (30,282) (37,305)	(43,319)
			(45,351) (49,331) (52,375)	(57,481)
			(65,395) (72,560) (79,409) (85,445) (89,461) (92,478) (95,483) (102,498)	(109,510)
			(112,515) (117,519) (123,522) (128,633) (132,573) (135,599) (142,581) (149,599) (151,601) (152,595) (156,585) (158,605) (159,609) (160,619)	
		};
		\addplot [
		color=red,
		mark=halfcircle*,
		] coordinates {(0,203) (5,209) (12,235) (14,240) (23,247) (30,252) (37,215)	(43,239)
			(45,248) (49,231) (52,275)	(57,281)
			(65,395) (72,290) (79,309) (85,345) (89,391) (92,278) (95,283) (102,298)	(109,310)
			(112,315) (117,319) (123,322) (128,333) (132,413) (135,399) (142,381) (149,299) (151,401) (152,425) (156,435) (158,505) (159,489) (160,519)	
		};
		\legend{$RND$,$SEQ$,$DHLB$,$DRALB$}
		\end{axis}
		
		\end{tikzpicture}
		\caption{Comparison of RND, SEQ, DHLB and DRALB in terms of Average Response Time when T=1000 and T$>$R}
		\label{Comparison of Time-Shared and Space-Shared resource allocation in terms of Failures}

		
	\end{figure}
	
	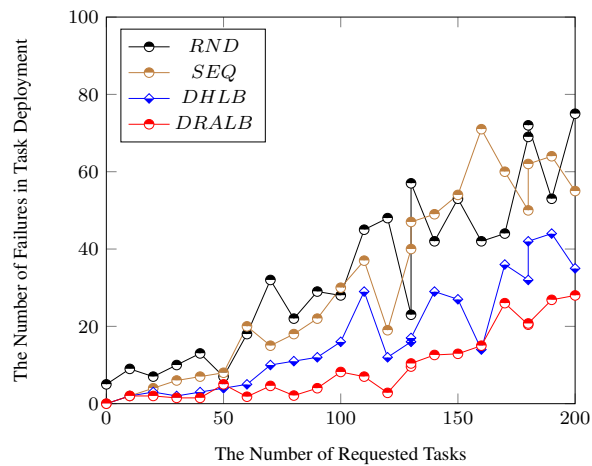
\begin{figure} [!h]
		\centering
		\begin{tikzpicture}[scale=0.9]
		\begin{axis}[
		xlabel={The Number of Requested Tasks},
		ylabel={The Number of Failures in Task Deployment},
		xmin=0, xmax=200,
		ymin=0, ymax=100,
		xtick={0,50,100,150,200},
		ytick={0,20,40,60,80,100},
		legend pos=north west,
		]
		\addplot 	[
		color=black,
		mark=halfcircle*,
		]
		coordinates  {
			(0,5) (10,9) (20,7) (30,10) (40,13) (50,7) (60,18) (70,32) (80,22) (90,29) (100,28) (110,45) (120,48) (130,23) (130,57) (140,42) (150,53) (160,42) (170,44) (180,69) (180,72) (190,53) (200,75)
		};
		\addplot 
		[
		color=brown,
		mark=halfcircle*,
		]  coordinates {
			(0,0) (10,2) (20,4) (30,6) (40,7) (50,8) (60,20) (70,15) (80,18) (90,22) (100,30) (110,37) (120,19) (130,40) (130,47) (140,49) (150,54) (160,71) (170,60) (180,50) (180,62) (190,64) (200,55)
		};
		\addplot
		[
		color=blue,
		mark=halfsquare*,
		] 
		coordinates {
			(0,0) (10,2) (20,3) (30,2) (40,3) (50,4) (60,5) (70,10) (80,11) (90,12) (100,16) (110,29) (120,12) (130,16) (130,17) (140,29) (150,27) (160,14) (170,36) (180,32) (180,42) (190,44) (200,35)
		};
		\addplot
		[
		color=red,
		mark=halfcircle*,
		]  coordinates {
			(0,0) (10,2) (20,2) (30,1.5) (40,1.5) (50,5) (60,1.8) (70,4.6) (80,2.1) (90,4) (100,8.2) (110,7) (120,2.8) (130,9.6) (130,10.4) (140,12.6) (150,12.9) (160,15) (170,26) (180,20.43) (180,20.75) (190,26.86) (200,28)
		};
		\legend{$RND$,$SEQ$,$DHLB$, $DRALB$}
		\end{axis}
		\end{tikzpicture}
		\caption{Comparision of RND, SEQ, DHLB and DRALB Resource Allocation in Failures}
		\label{Comparison of Time-Shared and Space-Shared resource allocation in terms of Failures}
		
	\end{figure}
	\begin{table} 
		\caption{Resource Utilization (in \%age)}\label{tbl:headings} 
		\centering
		\begin{tabular}
			{|p{1.5cm}||p{0.7cm}|p{0.7cm}|p{0.7cm}|p{0.9cm}||p{1.1cm}||p{0.7cm}|p{0.7cm}|p{0.7cm}|p{0.9cm}||p{1.1cm}||}
			\hline
			Allocation & \multicolumn{4}{c|}{When, T $<$ R} &Average & \multicolumn{4}{c|}{When, T $>$ R} & Average\\
			\cline{2-5}
			\cline{2-5}
			\cline{7-10}
			\cline{7-10}
			\cline{2-5}
			\cline{7-10}
			Policies & \multicolumn{4}{c|}{When, T $<$ R} & Wastage & \multicolumn{4}{c|}{When, T $>$ R} & Wastage\\
			\hline
			\hline
			\hline

			RND  &66.42  &69.61 &89.62 &70.33 &49.53\% &69.67  &78.72 &70.99 &85.68 &40.73\%\\

			SEQ &71.14  &78.41 &75.61 &80.52 &33.24\% &73.42  &76.14 &79.52 &80.42 &30.51\%\\
			
			DHLB &73.62  &77.52 &80.01 &76.21 &27.98\% &76.42  &79.32 &83.67 &79.42 &22.61\%\\
			
			DRALB &70.33 &79.39 &75.52 &77.67 &20.67\% &75.11  &80.12 &82.55 &81.52 &18.31\%\\
			\hline
		\end{tabular}
	\end{table}
	\begin{figure} [!h]
		\centering
		\begin{tikzpicture} [scale=0.9]
		\begin{axis}[
		xlabel={Traffic intensity},
		ylabel={Proportion of overloaded traffic},
		xmin=00, xmax=200,
		ymin=0, ymax=0.010,
		xtick={0,50,100,150,200},
		ytick={0,0.002,0.004,0.006,0.008,0.010},
		legend pos=north west,
		]
		\addplot	[
		color=black,
		mark=halfcircle*,
		]
		coordinates {
			(0,0.001) (0,0.0013) (10,0.002) (20,0.0024) (30,0.0029) (34,0.0035) (47,0.0043)	(54,0.0027)
			(65,0.0028) (69,0.0031) (70,0.0035)	(73,0.0037)
			(76,0.0040) (79,0.0041) (100,0.0045) (120,0.0046) (130,0.0048) (135,0.0049) (140,0.0054) (140,0.0052)	(140,0.0058)
			(142,0.0059) (147,0.0060) (149,0.0062) (150,0.0062) (150,0.0064) (150,0.0068) (150,0.0053) (164,0.0062) (166,0.0069) (176,0.0073) (177,0.0078) (187,0.008)
		};
		\addplot
		[
		color=brown,
		mark=halfcircle*,
		]  coordinates{
			(0,0.0005) (0,0.001) (10,0.0011) (20,0.0014) (30,0.0017) (34,0.0019) (47,0.0020)	(54,0.0029)
			(65,0.0038) (69,0.0039) (70,0.0039)	(73,0.0040)
			(76,0.0040) (79,0.0041) (100,0.0035) (120,0.0036) (130,0.0038) (135,0.0039) (140,0.0034) (140,0.0042)	(140,0.0048)
			(142,0.0049) (147,0.0050) (149,0.0052) (150,0.0051) (150,0.0052) (150,0.0058) (150,0.0043) (164,0.0052) (166,0.0049) (176,0.0053) (177,0.0058) (187,0.006)
		};
		\addplot [
		color=blue,
		mark=halfsquare*,
		] 
		coordinates{
			(3,0.0002) (6,0.0006) (10,0.0009) (20,0.001) (30,0.0012) (34,0.0013) (47,0.0015)	(54,0.0019)
			(65,0.0013) (69,0.0014) (70,0.0016)	(73,0.0016)
			(76,0.0018) (79,0.0019) (100,0.0019) (120,0.0020) (130,0.0021) (135,0.0022) (140,0.0023) (140,0.0024)	(140,0.0026)
			(142,0.0027) (147,0.0028) (149,0.0030) (150,0.0021) (150,0.0022) (150,0.0028) (150,0.0023) (164,0.0022) (166,0.0023) (176,0.0023) (177,0.0028) (187,0.003)
		};
		\addplot
		[
		color=red,
		mark=halfcircle*,
		]  coordinates{
			(3,0.0001) (6,0.0002) (10,0.0003) (20,0.0005) (30,0.0006) (34,0.0007) (47,0.0008)	(54,0.0009)
			(65,0.0013) (69,0.0009) (70,0.0002)	(73,0.0011)
			(76,0.0012) (79,0.0013) (100,0.0005) (120,0.0016) (130,0.0017) (135,0.0022) (140,0.0019) (140,0.0020)	(140,0.0016)
			(142,0.0017) (147,0.0018) (149,0.0024) (150,0.0027) (150,0.0022) (150,0.0022) (150,0.0019) (164,0.0018) (166,0.0023) (176,0.0021) (177,0.0024) (187,0.0029)
		};
		
		\legend{$RND$,$SEQ$,$DHLB$,$DRALB$}
		\end{axis}
		\end{tikzpicture}
		
		\caption{Proportion of overflow traffic}
		\label{Comparison of RND, SEQ and DRALB in proportion of overflow traffic}
	\end{figure}
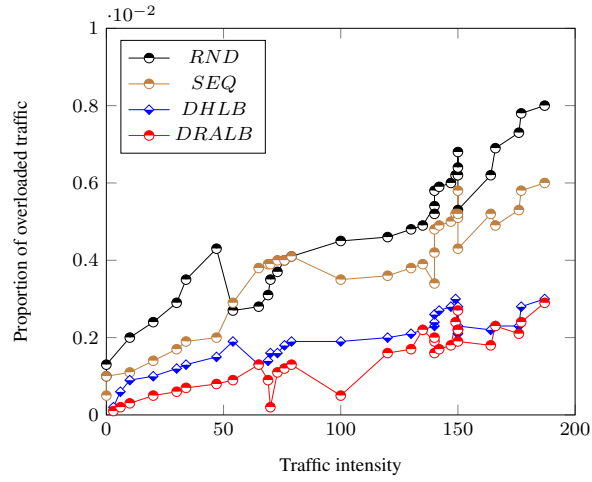
	\section{Conclusion}
	This paper presents an application SLA aware resource allocation scheme to analyse the resource requirements and allocate an appropriate number of physical machines for the particular deployment. This framework monitors the upcoming applications and demonstrate the SLA aware scheduling policy. The two performance parameters are focused to meet fairness constraints as per SLA violation while monitoring the applications. Experiments show that DRALB can improve effective load balancing in contrast to sequential, random placement and DHLB method. This SLA-aware cloud load balancer maximizes the benefit of cloud services, reduces the number of failures and minimizes the penalty function.  Performance evaluation demonstrates that DRALB achieves satisfactory outputs which reduces the wastage of resources and reduces the traffic upto 44.89\% and 58.49\% respectively for the experimental data while pointing out the observations of existing well-known algorithms.\\ 
	
	In future work, we will study the adaptive methods to better balance the tradeoff between
	SLA violation and the resource efficiency. Besides, adding more objectives into our model and
	then implementing the algorithms in a real cloud data centre constitute our future work.
	\\

	\section*{Acknowledgment}
	The authors would like a wonderful thanks to the DeitY (Department of Electronics and Information technology) for this research project. And we are extremely fortunate to get constant encouragement, support as well as appreciation.
	
\vskip 3pt plus -1fil

\begin{thebibliography}{}
		
		\bibitem{1}
		Yu, Rong, Yan Zhang, Stein Gjessing, Wenlong Xia, and Kun Yang.: 'Toward cloud-based vehicular networks with efficient resource management.' IEEE Network 27, no. 5, 2013: 48-55.	
		\bibitem{2}
		Chhabra Sakshi, and Ashutosh Kumar Singh.: 'A Probabilistic Model for Finding an Optimal Host Framework and Load Distribution in Cloud Environment.' Procedia Computer Science 125 (2018): 683-690.
		\bibitem{3}
		Dou, Wanchun, Xiaolong Xu, Xiang Liu, Laurence T. Yang, and Yiping Wen. "A Resource Co-Allocation method for load-balance scheduling over big data platforms." Future Generation Computer Systems 86 (2018): 1064-1075.	
		\bibitem{4}
		Tso, Fung Po, and Dimitrios P. Pezaros.: 'Improving data center network utilization using near-optimal traffic engineering.' IEEE transactions on parallel and distributed systems 24, no. 6 (2013): 1139-1148.	
		\bibitem{5}
		Liang, Quan, Jing Zhang, Yong-hui Zhang, and Jiu-mei Liang. "The placement method of resources and applications based on request prediction in cloud data center." Information Sciences 279 (2014): 735-745.	
		\bibitem{6}
		Ma, Teng, Jiangxing Wu, Yuxiang Hu, and Wanwei Huang.: 'Optimal VM placement for traffic scalability using Markov chain in cloud data centre networks.' Electronics Letters 53, no. 9 (2017): 602-604.	
		\bibitem{7}
		Zuo, Liyun, Shoubin Dong, Lei Shu, Chunsheng Zhu, and Guangjie Han. "A multiqueue interlacing peak scheduling method based on tasks classification in cloud computing." IEEE Systems Journal 12, no. 2 (2016): 1518-1530.
		\bibitem{8}
		Peng, Jun-jie, Xiao-fei Zhi, and Xiao-lan Xie. "Application type based resource allocation strategy in cloud environment." Microprocessors and Microsystems 47 (2016): 385-391.
		\bibitem{9}
		S. Chhabra   and  A.K. Singh.: 'Dynamic Hierarchical Load Balancing Model for Cloud Data Center Networks.' IET Digital Library, Volume 55, Issue 2, 24 January 2019, p. 94 – 96.
		\bibitem{10}
		Dou, Wanchun, Xiaolong Xu, Xiang Liu, Laurence T. Yang, and Yiping Wen. "A Resource Co-Allocation method for load-balance scheduling over big data platforms." Future Generation Computer Systems 86 (2018): 1064-1075.
		\bibitem{11}
		Alkhanak, Ehab Nabiel, and Sai Peck Lee. "A hyper-heuristic cost optimisation approach for Scientific Workflow Scheduling in cloud computing." Future Generation Computer Systems 86 (2018): 480-506.
		\bibitem{12}
		Zhao, Yangming, Yifan Huang, Kai Chen, Minlan Yu, Sheng Wang, and DongSheng Li. "Joint VM placement and topology optimization for traffic scalability in dynamic datacenter networks." Computer Networks 80 (2015): 109-123.	
		
	\end{thebibliography}
\end{document}